\documentclass[aps,twocolumn,prd,superscriptaddress,showpacs,nofootinbib]{revtex4-1}

\usepackage[utf8]{inputenc}

\usepackage{diagbox}

\usepackage{mathtools}
\usepackage{amsfonts}
\usepackage{mathrsfs}
\usepackage{bbm}
\usepackage{slashed}

\usepackage{graphicx}
\usepackage{color}
\usepackage{array}

\usepackage{placeins}
\usepackage{booktabs}
\usepackage{makecell}
\usepackage{epstopdf}
\usepackage[caption=false]{subfig}

\usepackage{xspace}
\usepackage{siunitx}
\usepackage{hyperref}
\usepackage[nameinlink]{cleveref}
\usepackage{appendix}

\usepackage{xifthen}
\usepackage{xcolor}
\hypersetup{
	colorlinks,
	linkcolor={red!75!black},
	citecolor={blue!75!black},
	urlcolor={blue!75!black}
}

\setkeys{Gin}{width=0.48\textwidth}

\graphicspath{
	{./figures/}
	{../figures/}
}

\def\s0#1#2{\mbox{\small{$ \frac{#1}{#2} $}}}
\def\0#1#2{\frac{#1}{#2}}

\newcommand{\gettitle}{Thermal properties of $\pi$ and $\rho$ meson}

\hypersetup{
	pdftitle={\gettitle},
	pdfauthor={Gao, Ding},
	pdfkeywords= {thermal property of hadron} {phase structure}  {Dyson-Schwinger equation},
	bookmarksopen=true,
	bookmarksopenlevel=2,
	bookmarksnumbered=true
}

\begin{document}

\title{\gettitle}

\author{Fei Gao}\thanks{gao@thphys.uni-heidelberg.de}
\affiliation{Institut f{\"u}r Theoretische Physik,
	Universit{\"a}t Heidelberg, Philosophenweg 16,
	69120 Heidelberg, Germany
}

\author{Minghui Ding}\thanks{mding.ectstar@gmail.com}
\affiliation{European Centre for Theoretical Studies in Nuclear Physics
and Related Areas (ECT$^\ast$) and Fondazione Bruno Kessler\\ Villa Tambosi, Strada delle Tabarelle 286, I-38123 Villazzano (TN) Italy\\
}

\begin{abstract}
We compute the pole masses and decay constants of $\pi$ and $\rho$ meson at finite temperature in the framework of Dyson-Schwinger equations and Bethe-Salpeter equations approach. Below transition temperature, pion pole mass increases monotonously, while $\rho$ meson seems to be temperature independent. Above transition temperature, pion mass approaches the free field limit of screening mass $\sim2\pi T$, whereas $\rho$ meson is about  twice as  large as  that limit. Pion and the longitudinal projection of $\rho$ meson decay constants have similar behaviour as the order parameter of chiral symmetry, whereas the transverse projection of $\rho$ meson decay constant rises monotonously as temperature increases.  The inflection point of decay constant and the chiral susceptibility get the same phase transition temperature. Though there is no access to the thermal width of mesons within this scheme, it is discussed by analyzing the Gell-Mann-Oakes-Renner (GMOR)  relation in medium.  These thermal properties of hadron observables will help us understand the QCD phases at finite temperature and can be employed to improve the experimental data analysis and  heavy ion collision simulations.
\end{abstract}
\maketitle
\section{Introduction}
\label{intro}
QCD phase structure at finite temperature  is with great interest of investigation both theoretically and experimentally. The investigations will lead to a thorough understanding of the matter formation and the universe evolution. The studies have suggested that the QCD matter experiences a crossover at zero chemical potential as the temperature increases.  At low temperature, the QCD matter could be well described by the hadron resonance gas, while it gradually becomes quark gluon plasma at high temperature~\cite{Luo:2017faz, Adamczyk:2017iwn, Andronic:2017pug, Stephanov:2007fk, Andersen:2014xxa,  Shuryak:2014zxa, Pawlowski:2014aha, Roberts:2000aa,Gao:2020qsj, Fischer:2018sdj, Fu:2019hdw, Braun:2019aow, Yin:2018ejt,Bellwied:2015rza,Bazavov:2018mes}.
  As the temperature changes, the thermal mass of hadrons will shift, the thermal width will usually get larger and  the decay of hadrons will then change.  In heavy quark sector, the production and dissociation of quarkonium can be regarded as the signal of the  existence of quark gluon plasma, and thus  settling down the thermal mass and decay  of quarkonium is important for heavy ion collisions  simulations~\cite{Brambilla:2010cs}. On the other hand, the chiral symmetry breaking or restoration in hot medium can be described by the thermal mass of the light meson and its relevant properties. The appearance of a turning point in the temperature dependence of thermal mass will give explanation to the occurrence of the chiral symmetry transition. Additionally, the deviation of  the light scalar resonance thermal mass from the mass in vacuum is relevant to the location of freeze out temperature~\cite{Bluhm:2020rha}. Therefore, it is with  desires  to illustrate the thermal properties directly within hadron observables.

  The thermal  hadron mass could be  separated into the  screening mass and pole mass owing to the $O(4)$ symmetry breaking at finite temperature. Screening mass, defined by the large distance behavior of hadron correlation function, is relatively easy to compute.  Studies on the light meson screening mass have been carried out with lattice QCD~\cite{Cheng:2010fe,Bazavov:2014cta,Ishii:2016dln,Bazavov:2019www} (see e.g.~\cite{Aarts:2017rrl} for an overview) and the functional approach~\cite{Blaschke:2000gd,Maris:2000ig,Wang:2013wk,Dorkin:2018aap}. It has been  predicted that the high temperature limit of meson screening mass is $M_{\rm scr}\sim 2\pi T$,  as it expected to be the propagation of the thermal quark~\cite{Eletsky:1988an,Florkowski:1993bq}.  Nevertheless, the relation between the screening mass and phenomenologically relevant observables is not clear, and hence,  it is   important to study the temperature dependence of the pole mass of hadrons.  Though the computation of pole mass in lattice QCD simulation usually encounters  the complicated temperature connection between the spectral function and the kernel in temporal correlation functions, there are some pioneering results for pole masses of baryons~\cite{Aarts:2017iai,Aarts:2018glk,Aarts:2019hrg}.

In the present work we employ the Dyson-Schwinger equations (DSEs) and Bethe-Salpeter equations (BSEs) in imaginary time formalism with Matsubara frequency to study the in medium properties of $\pi$ and $\rho$ meson, which essentially  characterize the  dynamical chiral symmetry breaking mechanism nonperturbatively. Even though the meson is not on shell on the Matsubara frequency, the eigenvalues of BSEs  on the Matsubara frequency could be employed to extract the pole mass and decay constants.  In employing the DSEs approach herein, we apply the quark gluon interaction   which can reproduce hadronic static properties at zero temperature. The interaction is extended at finite temperature via including the thermal mass from perturbative QCD computation. Within this scheme, we then obtain the pole masses and decay constants  for $\pi$ and $\rho$ meson in a large range of temperature. Moreover, by including the computation of the chiral condensate, we also have the opportunity to verify the  GMOR relation in medium.
Though the method is not sophisticated enough to extract the thermal width of meson since it is related to the imaginary part of the Bethe-Salpeter amplitudes which can not be directly obtained in the imaginary time formalism,  it can be argued that the finite thermal width would cause the deviation of GMOR relation at finite temperature, which is then analyzed in this work.

The remainder of this paper is organized as follows.
In Sec. \uppercase\expandafter{\romannumeral2} we reiterate briefly the DSEs and BSEs approach at finite temperature.  We highlight in this section how the pole mass is extracted from the BSEs in imaginary time formalism with Matsubara frequency.
  Sec. \uppercase\expandafter{\romannumeral3} contains our results of the temperature dependence of pole masses and decay constants of $\pi$ and $\rho$ meson, as well as the discussion on the GMOR relation at finite temperature.
Finally, we summarize in Sec.\uppercase\expandafter{\romannumeral4}.

\section{DSE-BSE scheme at finite temperature }

\subsection{Dyson-Schwinger Equations at finite temperature}

At finite temperature the quark propagator can be written as~\cite{Roberts:2000aa}
\begin{equation}\label{eq:quark2pointT}
  S^{-1}(p) =i \vec \gamma \cdot \vec p A(p)
 + i \gamma_0 \,p_0 C(p) +  B(p)\,,
\end{equation}
where $p=(\vec p, p_0=\omega_l)$ with $\omega_l=(2l+1)\pi T$ the fermion Matsubara frequency. The quark propagator satisfies the Dyson-Schwinger equation as

\begin{eqnarray}\label{eq:quark2pointTdse}
  S^{-1}(p)& =&Z^{\bot}_2 i
 \vec \gamma \cdot \vec p + Z^{\perp}_2 i \gamma_0 \, p_0+Z_4 m_{0}   +Z_1\Sigma(p) \,,\\
\Sigma(p) &=&  T\sum_n \! \int\frac{d^3{q}}{(2\pi)^3}\; {g^{2}} D_{\mu\nu} (p-q;T) \nonumber\\
 &&\times \frac{\lambda^a}{2} {\gamma_{\mu}} S(q) \frac{\lambda^a}{2}\Gamma_{\nu}\, ,
\end{eqnarray}

where $m_0$ is the current quark mass; $q=(\vec{q},\omega_n)$; $D_{\mu\nu}$ the gluon propagator; $\Gamma_{\nu}$, the quark-gluon vertex; $Z_{1 }(\zeta)$ and $Z_{4 }(\zeta)$ respectively, the vertex and mass renormalisation constants; $\zeta$ the renormalisation point; $Z^{\bot,\perp}_2 (\zeta)$, respectively, the spatial and  temporal quark wave function renormalisation constants. With this, the quark condensate can be defined as
\begin{equation}\label{eq:chiralcondG}
\langle\bar{q}q\rangle_T\simeq  -
T\sum_n \int \frac{d^3 q}{(2 \pi)^3}
tr \,S_{ } (q)\,,
\end{equation}
 The   light chiral condensate needs to be subtracted for  quarks with masses, which is then given by~\cite{Gao:2020qsj}
\begin{equation}\label{eq:chiralcondren}
\langle\bar{q}q\rangle =  \langle\bar{q}q\rangle_T-m_0\frac{\partial \langle\bar{q}q\rangle_T}{\partial m_0}\,.
\end{equation}

The quark DSE can be solved under a certain truncation of quark-gluon vertex, and also there will need a consistent truncation for the quark scattering kernel in BSEs.  The latter truncation entails careful consideration of axial vector/vector Ward identities in addition to the full computation of the three point correlation functions required in the formal one, and the truncations beyond the rainbow-ladder truncation have only been investigated at zero temperature~\cite{Chang:2009zb,Williams:2015cvx,Qin:2020jig}.      The rainbow-ladder truncation  is the first systematic, symmetry-preserving DSE truncation scheme which is accurate for ground-state vector- and isospin-nonzero-pseudoscalar-mesons owing to the corrections of these channels cancel via the Ward-Takahashi identities~\cite{Maris:2003vk,Ding:2019lwe,Ding:2019qlr}.
Therefore,   in addition to  the efforts of improving  the truncation of quark scattering kernel at finite temperature, here we focus on computing the properties of $\pi$ and $\rho$ meson with the rainbow-ladder truncation.

 The truncation scheme employs the tree level quark-gluon vertex with modeling the interaction kernel introduced in Ref.~\cite{Qin:2011dd} ,
$g^2D_{\mu\nu}(s)= {\cal P}_{\mu\nu}(k) {\cal G}(s=k^2)$ :
\begin{equation}\label{eq:gluonmodel}
 {\cal G}(s)=\frac{8\pi^2}{\omega^4}D e^{-s/\omega^2} +\frac{8\pi^2 \gamma_m \mathcal{F}(s)}{{\rm ln}[\tau+(1+s/\Lambda^2_{QCD})^2]} \, ,
\end{equation}
where: ${\cal P}_{\mu\nu}(k)=\delta_{\mu\nu}-\frac{k_{\mu}k_{\nu}}{k^2}$; $\gamma_m=12/(33-2N_{f})$, $N_{f}=4$, $\Lambda^{N_f=4}_{{\rm QCD}}=0.234\,$GeV; $\tau=e^2-1$; and $\mathcal{F}(s)=[1-\exp(-s/[4m_t^2])]/s$, $m_t=0.5\,$GeV. The interaction kernel involves a massive gluon propagator on the domain at $s=0$, which is consistent with that determined in recent studies of QCD's gauge sector~\cite{Bowman:2004jm,Cucchieri:2007ta,Boucaud:2010gr,Oliveira:2010xc,Fister:2011uw,Cyrol:2016tym, Cyrol:2017qkl,Gao:2017tkg,Aguilar:2008xm,Aguilar:2012rz,Aguilar:2015nqa,Aguilar:2019kxz,Aguilar:2019uob}. At finite temperature, the gluon propagator  separates into color-magnetic and electric  modes, i.e., the dimension corresponding to temperature will be isolated in order to allow the introduction of  $O(4)$ symmetry breaking~\cite{Roberts:2000aa}. We then have~\cite{Qin:2010nq,Gao:2017gvf}:
\begin{equation}
g^2 D_{\mu\nu}(\vec{k}, \Omega_{ln}) = {\cal P}_{\mu\nu}^{M} D_{M}(k) + {\cal P}_{\mu\nu}^{E} D_{E}(k)\,,
\end{equation}
where $\Omega_{ln}=\omega_l-\omega_n$, $\vec{k}=\vec{p}-\vec{q}$, $k=(\vec{k}, \Omega_{ln})$  and ${\cal P}_{\mu\nu}^{M,E}$ are, respectively, the color-magnetic and electric projection operators as:
\begin{eqnarray} \label{eq:GA2}
&&{\cal P}^M_{\mu\nu}(k)=(1-\delta_{0\mu})(1-\delta_{0\nu}) \left(\delta_{\mu\nu}-\frac{k_\mu k_\nu}{\vec{k}^2} \right)\,,\notag\\%
&&{\cal P}^E_{\mu\nu}(k)={\cal P}_{\mu\nu}(k)-{\cal P}^M_{\mu\nu}(k)\,,
\end{eqnarray}
  and
\begin{eqnarray}
{D_{M}(k)} &=&\mathcal{G}(k^2), \quad
{D_{E}(k)} =\mathcal{G}(k^2+{m_{g}^{2}}) \, ,
\end{eqnarray}
where  $m_{g}^{}$ is the thermal mass of the gluon and can be taken as $m_{g}^{2}=16/5T^2 $ according to   perturbative QCD calculations~\cite{Haque:2012my}.

\subsection{Bethe-Salpeter Equation in imaginary time formula}

The practical way of computing BSEs for mesons at finite temperature is through the imaginary time formula which is simply to change  the fourth component of all the  momentum in Euclidean space to Matsubara frequency~\cite{Blaschke:2000gd,Maris:2000ig,Wang:2013wk}.
 Applying the rainbow-ladder truncation, the homogeneous BSE at finite temperature can be described as:
\begin{eqnarray}\label{eq:genbse}
        \lambda(\vec{P}^2,\Omega^2_m)\Gamma^{ab}_{\pi,\rho}(k;P) &= & T\sum_n\int \frac{d^3 {q}}{(2\pi)^3}  g^2D_{\mu\nu}(k-q;T)\, \notag\\
         && \times\,\gamma_\mu\chi_{\pi,\rho}^{ab}(q;P)\gamma_\nu \,,
\end{eqnarray}
where

\mbox{$\chi_{\pi,\rho}^{ab}({q};P) := S(\vec{q}+\vec{P},\omega_n+\Omega_m) \Gamma_{\pi,\rho}^{ab}({q};P)
S (\vec{q},\omega_n)$} and $P=(\vec P,P_0=\Omega_m)$ with $\Omega_m=2m\pi T$. $\lambda(\vec{P}^2,P_0^2=\Omega^2_m)$ is the eigenvalue of the meson BSE.

The eigenvalue of the homogeneous BSE becomes $1$ when the meson propagator is on shell,  i.e., $$ \vec{P}^2+P_0^2+M(\vec{P}^2,P_0^2)=0\,,$$
where $M(\vec{P}^2,P_0^2)$ is the meson mass. On one hand, people could define the so called screening mass $M_{\rm scr}$ via putting $\Omega^2_m=0$, extending $\vec{P}$ into complex plane and then locating  the screening mass  at $\lambda(-M^2_{\rm scr},0)=1$~\cite{Cheng:2010fe,Bazavov:2014cta,Ishii:2016dln,Bazavov:2019www,Blaschke:2000gd,Maris:2000ig,Wang:2013wk}. On the other hand, the pole mass is in principle difficult to define since an analytic continuation  of the Matsubara frequency in the form of spectral representation is required, which is ~\cite{Bellac:2011kqa}:
 \begin{equation}
 \frac{1}{\Omega^2_m+\vec{P}^2+M(\vec{P}^2,\Omega_m^2)}=\int^{\infty}_{-\infty} d\omega \frac{\rho(\vec{P},\omega)}{\omega-i\Omega_m}\,.
 \end{equation}
 The pole mass is located at $\lambda(\vec{P}=0,P^2=-M_{\rm pole}^2)=1$ through replacing $i\Omega_m$  with  $M_{\rm pole}$ in the above spectral representation. Therefore, if people try to obtain the pole mass directly, the BSE in real time formula with the spectral representation is required. However, no matter how the formula is changed, the eigenvalue $\lambda(P^2)$ keeps to be an analytic function at least in a broad range of  $P^2\in[-M^2_{\rm pole}, \infty)$~\cite{Qin:2017lcd}. Therefore, one could proceed the way of constructing the  meson pole mass as  follows: We compute the  eigenvalues $\lambda(P^2=\Omega^2_m)$ at each $\Omega_m$ with $m=1,2,...,m_{\rm Max}$,  and extrapolate them  to obtain the pole mass of the meson $M_{\pi,\rho}$ at $\lambda(P^2=-M_{\pi,\rho}^2)=1$. The larger number of $m$ will certainly  lead  to  a  more precise extrapolation , and here we employ  $m_{\rm Max}=30$ practically.

 The decay constant  of meson also splits into temporal and spatial components at finite temperature. Here with the definition of momentum $P=(0,\Omega_m)$, the decay constant is then projected to  the temporal part in consistency with the pole mass. The spatial component of decay constant could be obtained at the location of screening mass for Bethe-Salpeter equation with momentum $P=(\vec{P},0)$.
\subsubsection{$\pi$ meson}
The essential case of interest is the temperature dependent behaviour of pion, which is the simplest two-body system as well as the Goldstone mode of QCD~\cite{Maris:1997hd}. The Bethe-Salpeter amplitude of pion outlined in Eq.(\ref{eq:genbse}) is of the general form
\begin{eqnarray}\label{eq:pionbsa}
 \Gamma_{\pi}(q;P)=i\gamma_5\tau^\pi_1(q;P)+\gamma_5 P\!\!\!\!/\ \tau^\pi_2(q;P).
 \end{eqnarray}
The computation in vacuum   shows that these two components are dominant while the other two components contribute around $5\%$ to the mass and decay constant. We then drop the other two terms with higher order Lorentz structures, which is the  most practical choice of theoretical studies on hadron phenomenological observables at finite temperature.  We  limit ourselves to this case, and further investigations with the complete set of Dirac basis can be the supplement of this work.

The decay constant of pion can also be extrapolated from $P^2=\Omega^2_m$ to $P^2=-M_\pi^2$ after the mass  is located. The definition of pion decay constant  is given as:
\begin{eqnarray}\label{eq:piondecay}
 f_{\pi}(P^2)= \frac{Z_2}{P^2} T\sum_n\int \frac{d^3 {q}}{(2\pi)^3}{\rm tr}\left[i\gamma_5P\!\!\!\!/\ \chi_{\pi}({q};\Omega_m)\right]\,,
 \end{eqnarray}
 which is the residue at the pion pole in the axial-vector vertex~\cite{Maris:1997hd,Maris:2003vk}.

 By projecting pion Bethe-Salpeter wave function  onto $\gamma_5$ channel, we could also define a quantity related to quark condensate, which is
 \begin{eqnarray}\label{eq:pionrho}
ir_{\pi}(P^2)= Z_4T \sum_n\int \frac{d^3 {q}}{(2\pi)^3}{\rm tr}\left[i\gamma_5\chi_{\pi}({q};\Omega_m)\right]\,.
 \end{eqnarray}

 In particular, the preservation of the axial-vector Ward-Green-Takahashi identity at zero temperature yields the mass relation~\cite{Maris:1997hd}
\begin{eqnarray}
\label{eq:masr}
f_\pi M^2_\pi=2m(\zeta)r_\pi(\zeta)\,.
\end{eqnarray}
The quantity $r_\pi$ is related to quark condensate in chiral limit with
 \begin{eqnarray}
 \lim_{m\rightarrow 0}r_\pi(\zeta)=\frac{\langle\bar{q}q\rangle^0}{f^0_\pi}\,,
 \end{eqnarray}
 where $\langle\bar{q}q\rangle^0$ is the chiral condensate; $f^0_\pi$ the pion decay constant in chiral limit.
 It indicates that the mass relation is equivalent to the GMOR relation:
 \begin{eqnarray}\label{eq:gmor}
f^2_\pi M^2_\pi=2m(\zeta)\langle\bar{q}q\rangle\,.
\end{eqnarray}

The mass relation in Eq.(\ref{eq:masr}) and/or  the related GMOR relation could be derived from axial vector Ward identity by putting  pion on-shell, therefore, such relations could be connected to the thermal width of pion. It is then interesting to check the behaviour of GMOR relation at finite temperature.

\subsubsection{$\rho$ meson}

The other case of great interest is the $\rho$ meson, with its Bethe-Salpeter amplitude outlined in Eq.(\ref{eq:genbse}) takes the general form~\cite{Maris:1999nt}
 \begin{eqnarray}\label{eq:rhobsa}
   \Gamma_{\mu,\rho}(q;P)=i\gamma^T_\mu \tau^\rho_1(q;P)+ q^T_\mu \tau^\rho_2(q;P),
  \end{eqnarray}
  with $F^T_\mu=\mathcal{P}_{\mu\nu}F_\nu$.  Here we practically consider two  Lorentz structures for $\rho$ meson, which are the dominant two terms while in principle there are eight Lorentz structures in the complete set of the vector Bethe-Salpeter amplitude ~\cite{Maris:1999nt}. Besides that, if trying to reflect the impact of $O(4)$ symmetry breaking, people need to split $\gamma^T_\mu$ and $q^T_\mu$ into their longitudinal  and transversal modes~\cite{Blaschke:2000gd}. Consequently, one shall see distinguishing temperature dependence of the transversal  $\rho$ meson from the longitudinal one. Instead of doing that, we keep their original form in the Bethe-Salpeter amplitude, leaving that possibility for further investigation. However, noticing that in the Bethe-Salpeter equations,  the dominant contribution of such temperature induced splitting is from  the gluon and  quark propagator  as we have considered here,   the present approximation that keeps $O(4)$  symmetry for the Lorentz structures of vertex  is acceptable.

It is also straightforward to consider the decay constants of $\rho$ meson, and they are
\begin{eqnarray}\label{eq:rhodecay}
f_{\rho}(P^2) &=&\frac{Z_2}{3\Omega_m} T\sum_n\int \frac{d^3 {q}}{(2\pi)^3}{\rm tr}\left[i\gamma_\lambda\chi_\lambda({q};\Omega_m)\right]\,,\\
f^T_{\rho}(P^2) &=& \frac{Z_T }{3P^2} T\sum_n\int \frac{d^3 {q}}{(2\pi)^3}{\rm tr}\left[i\sigma_{\mu\lambda}P_\mu\chi_{\lambda}({q};\Omega_m)\right]\,,\notag
\end{eqnarray}
with $Z_2$  is  the quark wave function renormalisation constant and $Z_T$  is  the renormalisation constant for the tensor vertex. These  two decay constants are both gauge- and Poincar\'e-invariant, but $f_\rho^T$ is renormalisation scale dependent~\cite{Gao:2014bca}.

\section{Numerical results}

We first fix all the needed parameters through matching pion properties in vacuum with experimental data. We set the renormalisation scale at $\zeta=19$ GeV,  which is the typical  choice in a bulk of extant studies~\cite{Maris:1997tm,Maris:1999nt}. The  parameter of interaction in Eq.(\ref{eq:gluonmodel}) is taken as $D\omega=(0.8\,\rm GeV)^3$ and $\omega=0.5$ GeV, and one can expect computed observables to be practically insensitive to the choice of $D$ or $\omega$ on a reasonable domain  with keeping the interaction strength $D\omega$ unchanged~\cite{Qin:2011xq}. The light current quark mass is $m(\zeta=19\,\rm GeV)=3.4~{\rm MeV}$, corresponding to the renormalisation-group-invariant mass $\hat{m}=6~{\rm MeV}$. Then we can get the vacuum property of $\pi$ and $\rho$ meson as follows:
\begin{eqnarray}
&&m_\pi = 138\,\textrm{MeV}\,,\quad   f_\pi=97 \,\textrm{MeV}\,, \\
&&m_\rho = 760\,\textrm{MeV}\,,\quad   f_\rho=153\,\textrm{MeV} \,,\quad   f^T_\rho=110\,\textrm{MeV}.\notag
\end{eqnarray}
With this, we then extend the computation into finite temperature region as analyzed above.

\subsection{Pole masses of $\pi$ and $\rho$ meson}

 As mentioned above, we computed the eigenvalue of BSE at Matsubara frequency $P_0=2 m \pi T$ and then extrapolate it to $\lambda(P_0^2=-M_{\text{pole}}^2)=1$. This method can be analog to the imaginary chemical potential approach  of lattice QCD, where the phase transition line is extrapolated through the information in the chemical potential region \cite{deForcrand:2002hgr,deForcrand:2010he,Philipsen:2016hkv,Bellwied:2015rza,DElia:2016jqh,Borsanyi:2018grb,Aarts:2013uxa}. Here  the eigenvalue of BSEs shares similar character to the phase transition line,  since  we know the eigenvalue increases smoothly and monotonously from $0$ to $1$  as  $P_0^2$ approaches $-M^2_{\text{pole}}$ from $P_0^2=\infty$,  it can be safely extrapolated from positive to negative $P_0^2$ till around $P_0^2\sim-M^2_{\text{pole}}$.

\begin{figure}[t]
\centerline{\includegraphics[width=0.50\textwidth]{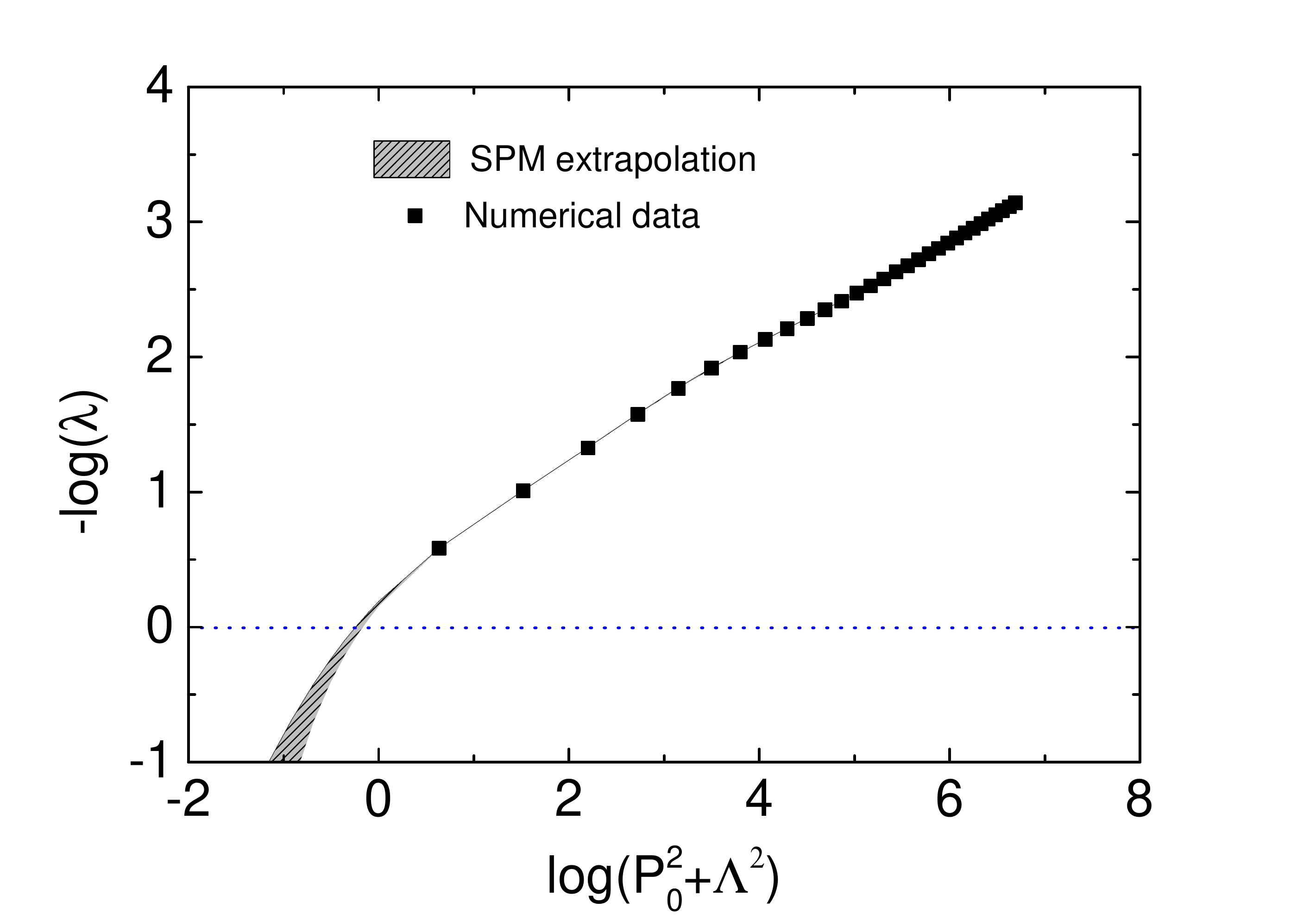}}
\caption{The extrapolation for eigenvalue of pion BSE at $T=150$ MeV with $\Lambda=1$ GeV. The pole mass is obtained at the location $\lambda(P_0^2=-M^2_{\text{pole}})=1$.} \label{fig:fitM}
\end{figure}

 In detail, we herein applied the Schlessinger Points  Method (SPM) of extrapolation based on the work by Thiele and Schlessinger \cite{PhysRevLett.16.1173,PhysRev.167.1411} and employed recently in a number of publications, see e.g. \cite{Tripolt:2016cya,Tripolt:2018xeo,Binosi:2019ecz,Santowsky:2020pwd}. The results are fairly stable with respect to different choices of input regions.  After performing the procedure with random-selected data points, one could also give a statistical error estimate. Additionally, in order to get a more stable extrapolation, we change the variables $P_0^2$ and $\lambda(P_0^2)$ into $x={\rm log}(P_0^2+\Lambda^2)$ with $\Lambda$ being roughly chosen at the order of  $ M_{\rm pole}(T)$ and $y=-{\rm log}(\lambda)$. Then we apply the SPM  based on the dataset $y_m=f(x_m)$. As an example, we depict the extrapolated eigenvalue of pion BSE at $T=150$ MeV in Fig.~\ref{fig:fitM}, and the band is given by randomly selecting the data sets and running the extrapolation as described above.  The pole mass is obtained at the location $\lambda(P_0^2=-M^2_{\text{pole}})=1$. The extrapolation can be done at each temperature  following this procedure, and the pole mass at finite temperature  is then obtained.

\begin{figure}[t]
\centerline{\includegraphics[width=0.50\textwidth]{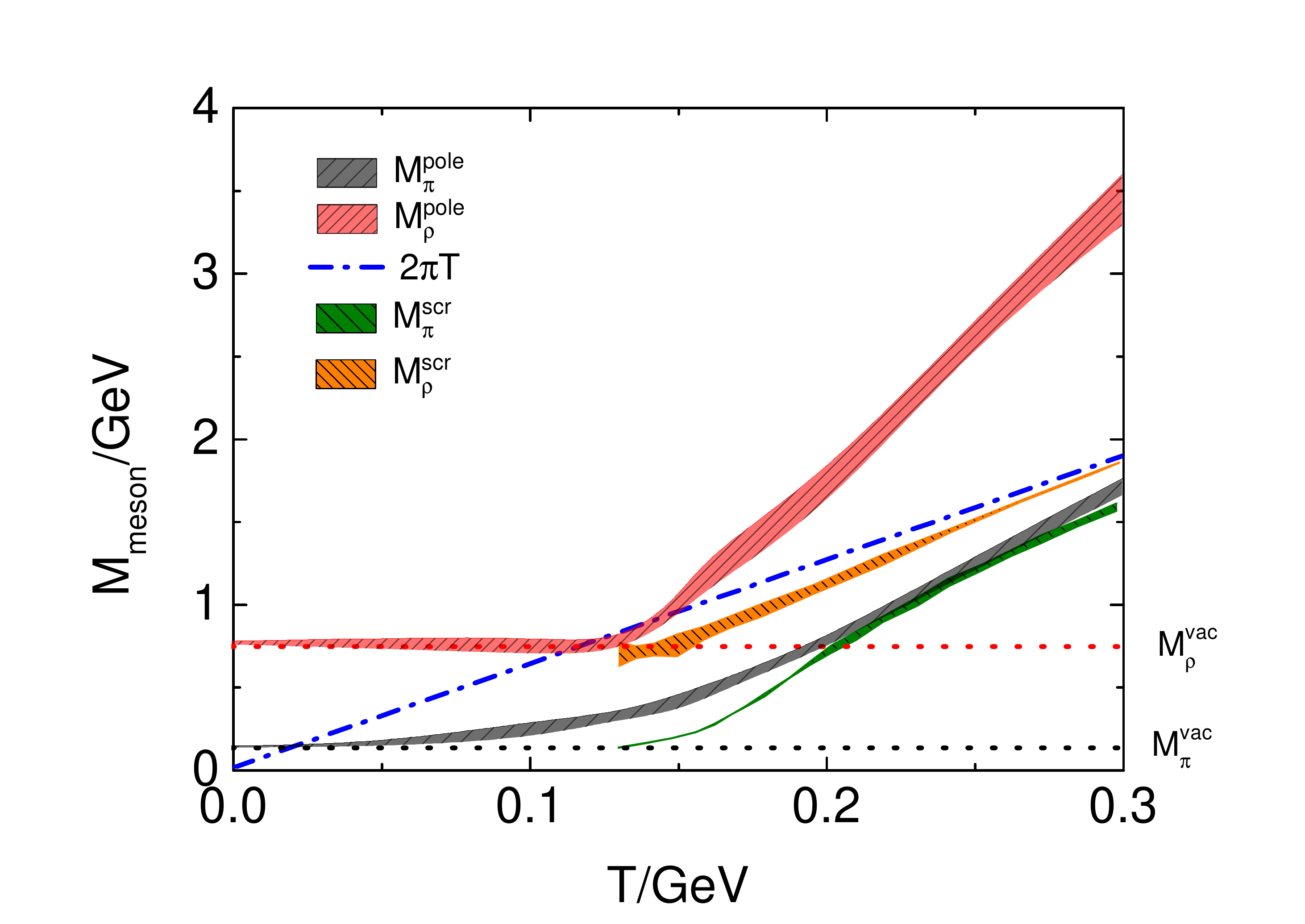}}
\caption{Pole masses of $\pi$ (\emph{${\color{black} black}$}) and $\rho$ (\emph{${\color{black} red}$}) meson at finite temperature as well as the free field limit of screening mass (\emph{dash-dotted curve}) $M_{{\rm scr}}\sim2\pi T$.  The bands mark the domain of the results obtained using SPM extrapolation. The $\pi$    and $\rho$  mass   in vacuum, $M^{vac}_{\pi}$ and  $M^{vac}_{\rho}$,  are  included to guide comparison.   The screening mass for $\pi$ (\emph{${\color{black} olive}$}) and $\rho$ (\emph{${\color{black} orange}$}) mesons are from lattice QCD (lQCD)~\cite{Bazavov:2019www}. } \label{fig:mass}
\end{figure}

We then compute the $\pi$ and $\rho$ meson pole masses at different temperature as shown in Fig.~\ref{fig:mass}. As the temperature increases, the mass of pion increases monotonously, while the mass of $\rho$ meson has no significant shift only decreasing slightly around $2\%$ till $T_{s}=0.12$ GeV and then rises rapidly. The weak $T$-dependence behaviour of $\pi$ and $\rho$ meson pole  masses at small temperature is qualitatively consistent with the response of screening mass to temperature as in DSE approach \cite{Blaschke:2000gd,Maris:2000ig,Wang:2013wk}. Below transition temperature in the hadronic phase, where the chiral symmetry is dynamically broken, it  results in a relatively stable pattern of the ground states in both pseudoscalar and vector channel and then the pole mass rapidly increases after phase transition.   In the region of phase transition, the smooth behaviour of  $\pi$ and $\rho$ meson pole masses with respective to $T$ also indicates a crossover rather than a phase transition.

 In detail, we  compare our obtained $\pi$ and $\rho$ meson pole masses with the screening mass results from lattice QCD simulation~\cite{Bazavov:2019www}. At low temperature, the pole mass of pion differs from the screening mass. The screening mass in lattice QCD simulation has reached   the vacuum value at relatively high temperature, while our pole mass increases monotonously from its zero temperature value, and is consequently larger than $M_{\pi}^{vac}$ at $T_s$. For $\rho$ meson,  though without  data at lower temperature, the screening mass  from lattice QCD simulation at $T_s$  has reached  the vacuum value $M_{\rho}^{vac}$. This feature is consistent with our $\rho$ meson pole mass, which has no significant shift till $T_s$ and then rises rapidly as temperature increases.

At high temperature, the screening mass of pion from lattice QCD simulation and the obtained pole mass here agree well with each other   around $T\sim0.2$ GeV, and both masses gradually reach the free field limit. They remain  around $12\%$ smaller than the limit at $T\sim0.3$ GeV, and this discrepancy reveals the strong coupled property of QCD matter at the temperature nearly above phase transition temperature. It is also worth mentioning  that  the screening mass of pion  will remain smaller at very large temperature compared to the computation in the weak coupling picture~\cite{Bazavov:2019www}.  As for $\rho$ meson,   the pole mass of pion gets close to the free field limit above the critical temperature, while the $\rho$ meson pole mass is twice as large as this limit.  With the caveat of the present truncation of Bethe-Salpeter equation, this  is consistent with  the description of  $\rho$ meson as a $\pi-\pi$ resonance state. The  screening mass of $\rho$ meson gets close to the free field limit above the critical temperature $T_c$. Though the discrepancy of the pole masses here between pseudoscalar and vector meson is qualitatively different from that of the screening masses, it has been shown that the  screening mass of $\pi$ meson is also notably  smaller than $\rho$ meson screening mass till $T\sim1$ GeV in lattice QCD simulation~\cite{Bazavov:2019www},  which reveals that it  still remains considerable non-perturbative effect of QCD on the thermal properties associated with bound states.

\subsection{Decay constants of $\pi$ and $\rho$ meson}

\begin{figure}[t]
\centerline{\includegraphics[width=0.50\textwidth]{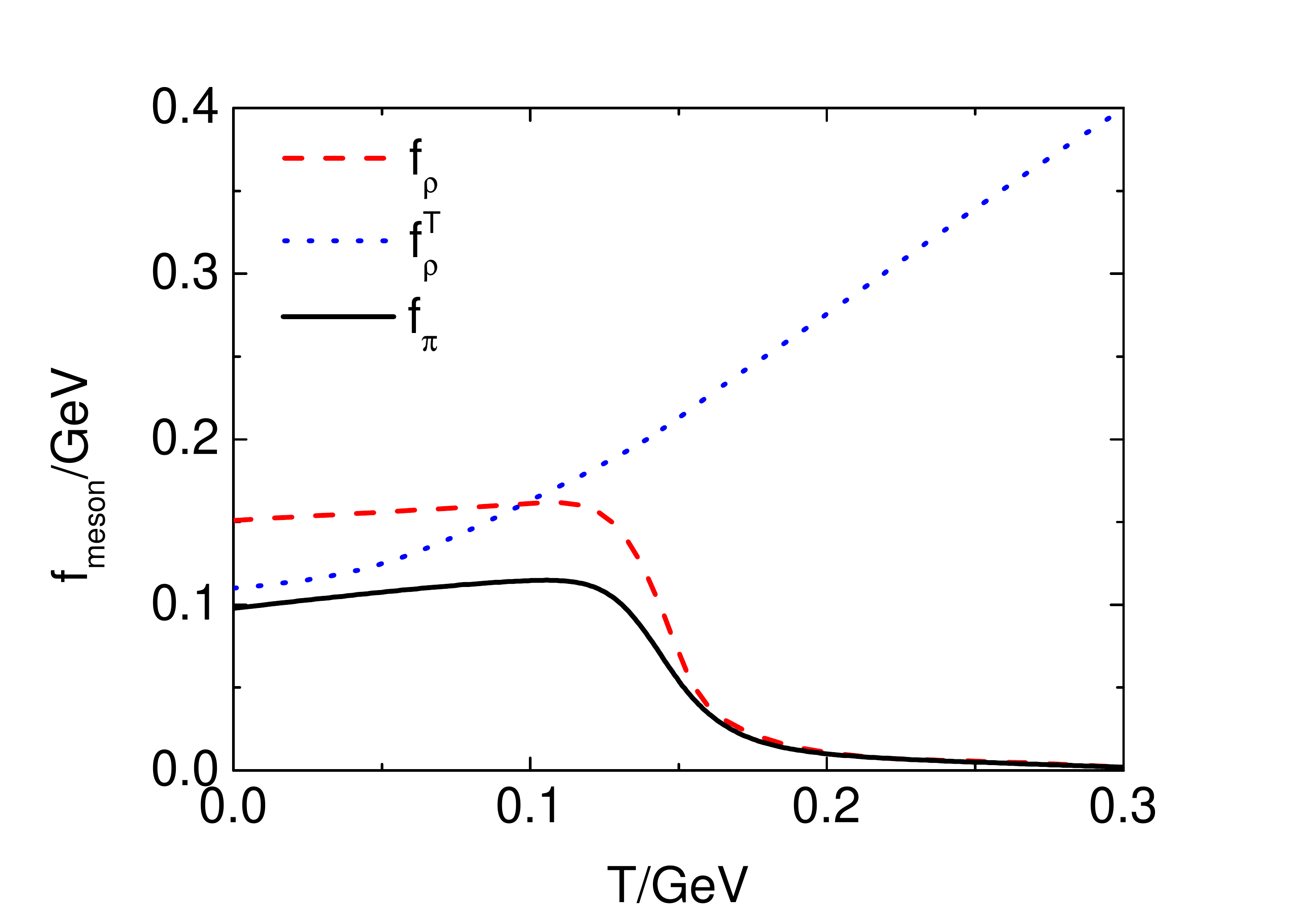}}
\caption{Decay constants of pion, $f_{\pi}$ (\emph{solid curve}) in Eq.(\ref{eq:piondecay}) and $\rho$ meson, $f_\rho$ (\emph{dashed curve}) and $f^T_\rho$ (\emph{dotted curve}) in Eq.(\ref{eq:rhodecay}) at finite temperature.} \label{fig:decay}
\end{figure}
Hitherto we have canvassed $\pi$ and $\rho$ meson thermal pole masses, it is also important in understanding their corresponding decay properties.    The extrapolation for decay constants needs the location of meson mass, and here we take the central value of the extrapolation for meson mass above.  The temperature dependence of $\pi$ and $\rho$ meson decay constants is illustrated in Fig.~\ref{fig:decay}. As the temperature increases, the decay constant of pion goes up very slightly till around $T=T_s$ and then declines rapidly to zero. Noticing that the light quark dynamical mass function in the quark propagator,  is also almost $T$-independent below a critical temperature, and then goes to zero. It should not be surprising of this resembling behaviour since the Bethe-Salpeter amplitude of pseudoscalar meson could be directly related  to  the quark mass function via the Ward identity~\cite{Maris:1997hd}. It is evident that both pion decay constant and the light quark dynamical mass function are equivalent order parameters for chiral symmetry restoration.  Below transition temperature, chiral symmetry is broken, and its order parameters  become nonzero. Above $T_c$, chiral symmetry get restored, and order parameters vanish quickly. Compared to other studies,  temperature dependence of pion decay constant here is qualitatively consistent with DSE results~\cite{Maris:2000ig,Wang:2013wk} and lattice QCD simulation~\cite{Laermann:2012sr}.

\begin{figure}[t]
\vspace{-0.33cm}
\centerline{\includegraphics[width=0.52\textwidth]{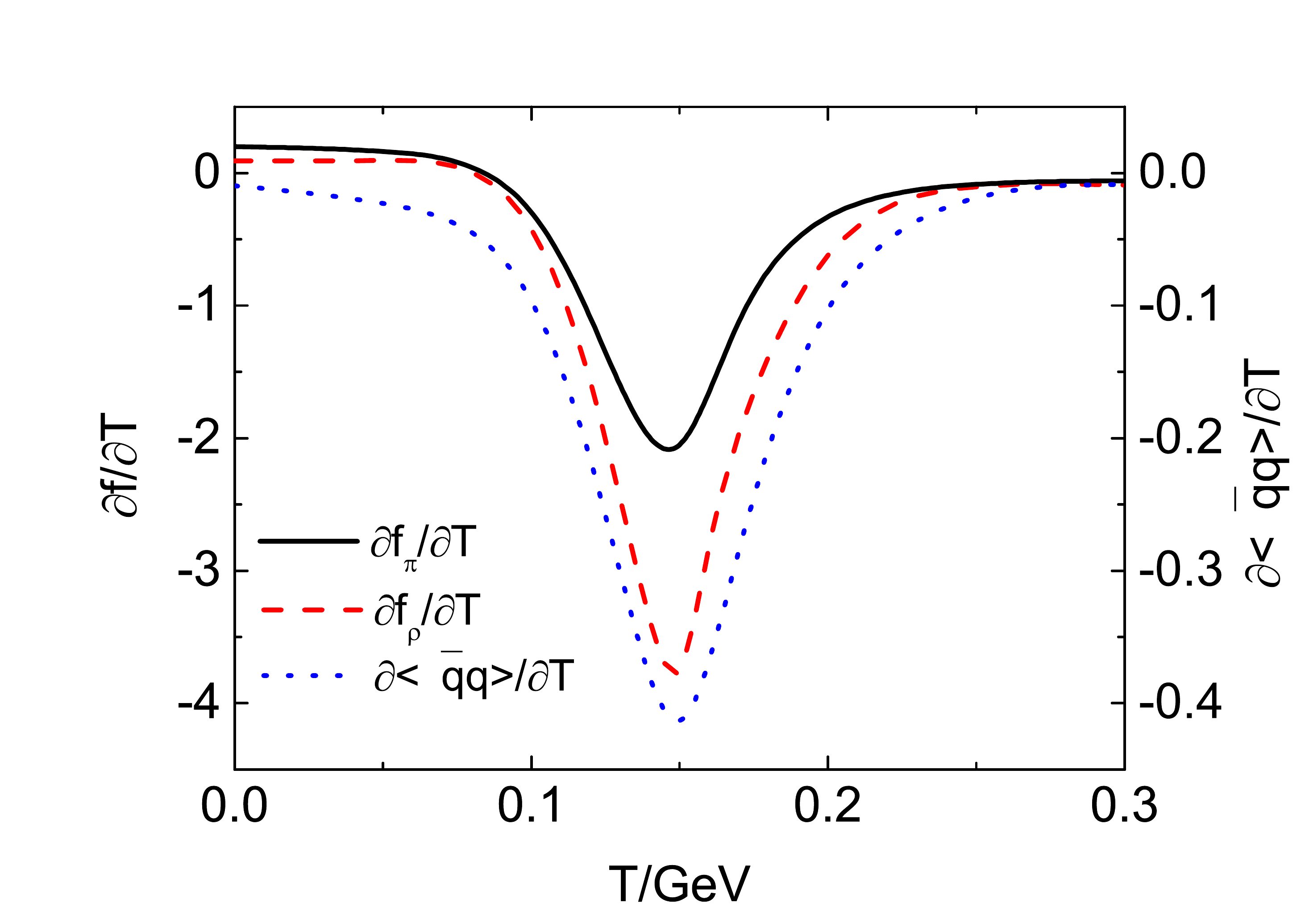}}
\caption{Temperature derivative of decay constants $f_{\pi}$ (\emph{solid curve}) and $f_\rho$ (\emph{dashed curve}), along with quark chiral susceptibility $\chi=\partial \langle \bar{q}q\rangle /\partial T$ (\emph{dotted curve}).} \label{fig:sus}
\end{figure}
 The longitudinal decay constant of $\rho$ meson has similar behaviour as pion's.    It slightly depends on temperature within the hadronic phase, and then chiral symmetry is rapidly restored above the transition temperature, apart from the explicit symmetry breaking by the current quark mass. The transverse decay constant behaves completely different however, which rises monotonously as temperature increases. The ratio of $f_\rho/f^T_\rho$ can be related to the proportion of $S-$ and $D-$ wave content of  the $\rho$ meson~\cite{Gao:2014bca}. Therefore, considering the behaviour of two decay constants, as the temperature increases,  one would find that  the $D-$ wave contribution becomes larger. Additionally, the higher order Lorentz structures in the $\rho$  meson Bethe-Salpeter amplitude could play an important role in computing an accurate value for $f^T_{\rho}$ at finite temperature, because they contain the detailed contributions of angular momentum.

Noticing that the decay constants $f_\pi$ and $f_\rho$ own similar behaviour as the order parameter, quark condensate, we then compare the temperature derivative of the decay constants with the chiral susceptibility, defined by the temperature derivative of quark condensate, i.e., $\chi=\partial \langle \bar{q}q\rangle/\partial T$. In Fig.~\ref{fig:sus} we can see the inflection point of pion decay constant, i.e., $\partial^2 f_\pi/\partial T^2=0$ almost coincides with that of $\rho$ meson decay constant.  In detail, the transition temperature associated with the inflection point of pion decay constant is $T^{f_\pi}_c=146~{\rm MeV}$, and that of $\rho$ meson is $T^{f_\rho}_c=149~{\rm MeV}$ compared to that  determined  by the inflection point of  quark condensate as $T^{\langle\bar{q}q\rangle}_c=150~{\rm MeV}$~\cite{Gao:2017gvf}. On average, our estimate is
\begin{equation}
	 T_c=(148\pm2)\ {\rm MeV}\,.
\end{equation}
 It is consistent with the chiral phase transition temperature from functional methods~\cite{Fischer:2014ata,Gao:2020qsj,Fu:2019hdw} and also lattice QCD which is in a range from $147$ to $165\ {\rm MeV}$ in Ref.~\cite{Aoki:2009sc,Borsanyi:2010bp} and $T^{lQCD}_c=(154\pm9)\ {\rm MeV}$ in Ref.~\cite{Bazavov:2011nk}.

In general, the decay constants could be regarded as a criterion of chiral transition. The  fact that chiral phase transition temperature defined with the temperature dependence of $\pi$ and $\rho$ meson decay constants and from the chiral condensate  coincide can be viewed as a direct evidence of the chiral phase transition from the physical observables.

\subsection{GMOR relation at finite temperature}

\begin{figure}[t]
\centerline{\includegraphics[width=0.50\textwidth]{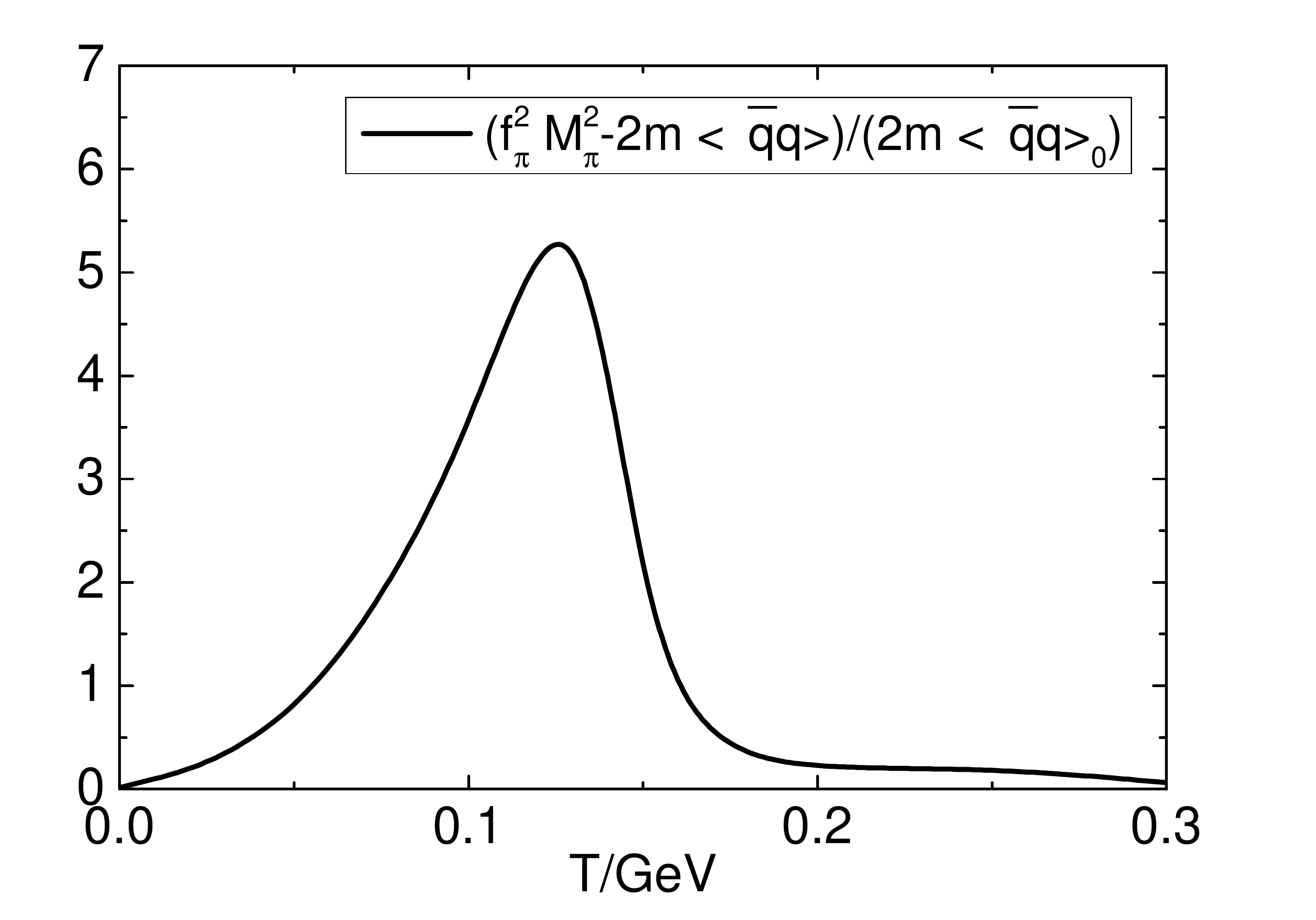}}
\caption{Deviation of GMOR relation at finite temperature, normalized by the quark condensate product in vacuum.} \label{fig:gor}
\end{figure}

 GMOR relation as in Eq.(\ref{eq:gmor}) can be derived by putting the axial vector Ward identity on mass shell of pion. The derivation of GMOR relation will rely on the assumption of a pole structure for pion. It has been argued that the axial vector Ward identity still holds at finite temperature~\cite{Pisarski:1996mt,Maris:2000ig,Hou:1997vb}, and hence the deviation of GMOR relation  indicates a finite thermal width of pion, which then drives pion spectral  function  away from a pole structure.  The  Ward identity  at finite temperature will certainly lead to some relation similar to GMOR relation, but the relation will deviate against the vacuum formula with additional temperature correction terms~\cite{Fu:2009zs,Dominguez:1996kf}.

 The deviation of the GMOR relation is shown in Fig.~\ref{fig:gor}.  Strictly speaking,  the GMOR relation at zero temperature is only satisfied with all four Lorentz structures of pion Bethe-Salpeter amplitude, however, the other two components in addition to the dominant  two structures computed here in Eq.~(\ref{eq:pionbsa}) barely have impact on the pion mass and decay constant. With only taking into account of  the two dominant components in Eq.~(\ref{eq:pionbsa}), the GMOR relation is well preserved with a small deviation less than $4\%$.  The deviation in vacuum is negligible, and then,   a  clearly remarkable increase  of the deviation has emerged when $T$ is approaching the critical temperature, and  above $T_c$, it vanishes drastically.   The experiments indicate that the matter near the phase transition is in a strongly-coupled  state, and the ratio of shear viscosity and entropy density is nearly the lower bound at phase transition point \cite{Sharma:2009zt,Majumder:2007zh,Song:2010mg}. The deviation of GMOR relation can be  then regarded  as  a signal for this strongly-coupled property of the matter in the phase transition region, where the thermal width of pion is  generated via Landau damping mechanism~\cite{Brambilla:2010cs,Qin:2013aaa,Qin:2014dqa}.   Moreover, the non monotonous behaviour  exhibits the change of  the  pion structure  during the phase transition, and the rapid decrease after phase transition also  indicates  the dissociation of pion.

\section{Summary}

In this work, the hadronic observables at finite temperature have been studied in the framework of DSEs and BSEs approach. As the temperature increases, the pole mass of pion becomes larger  monotonously and after the chiral phase transition, the pole mass gradually reaches the same limit as screening mass, $M_{{\rm scr}}\sim 2\pi T$ at high temperature. The pole mass of $\rho$ meson is quite stable till $T\sim 0.8 T_c$, and then rapidly grows at high temperature.  The mass of  $\rho$ meson   approaches  twice as large as  pion's at high temperature, which is consistent with picture of  the $\rho$ meson as the resonance of two pions.

After obtaining the location of the masses,  we compute the decay constants of $\pi$ and $\rho$ mesons. The decay constant of pion and the longitudinal decay constant of $\rho$ meson show similar behaviour as a function of temperature. In the hadronic phase, these quantities are barely dependent on temperature, and then drop  rapidly in the phase transition region. Thus, the decay constant is strongly related to the phase transition and can be employed as the criterion of chiral phase transition. They give the consistent chiral phase transition temperature as the quark condensate. The transversal decay constant of $\rho$ meson here shows a monotonously increasing behaviour as temperature increases.

Even though this method cannot directly give the information of the thermal width of mesons, the strong deviation of GMOR relation  indicates the strongly-coupled property of QCD matter in the phase transition region.  The non monotonous behaviour also exhibits the change of  the internal structure of mesons during the phase transition.

A straightforward and worthwhile extension of this work is the consideration in the pole masses of the $\sigma$ and $a_1$ meson. The proper calculation of the scalar and the axial-vector channel is complicated even at zero temperature, since one must include other Lorentz structures in quark-gluon vertex beyond rainbow-ladder approximation  in order to give a correct description of the angular momentum. Despite of this, the research on the temperature dependence of scalar and the axial-vector channel will   nevertheless provide us insights  into the difference of  parity partners. The other possible extension is to consider the mesons with strange quark and also quarkonium. It has been brought out by lattice QCD simulation that the screening masses of meson including strange quark will  give a  higher $T_c$~\cite{Bazavov:2014cta,Bazavov:2019www}. For the quakonium, the $J/\Psi$  production  is especially important for helping   understand experimental data of heavy ion collisions~\cite{Rapp:2009my,Yao:2018zrg}. Therefore, it will be of high value to extend the computation to these observables.
\begin{acknowledgements}
We thank Jan M. Pawlowski, Joannis Papavassiliou, Craig D. Roberts, Daniele Binosi, Lei Chang and Sixue Qin for discussions. F.~Gao is supported by the Alexander von Humboldt foundation.
\end{acknowledgements}
\bibliography{reflib}
\end{document}